\documentclass[aps,preprint,showpacs,floatfix]{revtex4}
\usepackage{graphicx,bm,amsmath,dcolumn}
\usepackage{epsfig}
\graphicspath{{figures}}
\begin{document}

\title{The two-point resistance of a resistor network: \\ A new formulation
and application to the cobweb network}

\author{N.Sh. Izmailian}
\email{izmail@yerphi.am; ab5223@coventry.ac.uk}
\affiliation{Applied Mathematics Research Center, Coventry University, Coventry CV1 5FB, UK}
\affiliation{Yerevan Physics Institute, Alikhanian Brothers 2, 375036 Yerevan, Armenia}

\author{R. Kenna}
\email{R.Kenna@coventry.ac.uk}
\affiliation{Applied Mathematics Research Center, Coventry University, Coventry CV1 5FB, UK}

\author{F.Y. Wu}
\email{fywu@neu.edu}
\affiliation{Department of Physics, Northeastern University, Boston, MA 02115, USA}

\date{\today}

\begin{abstract}
We consider the problem of  two-point resistance  in a resistor
network previously  studied  by one of us [F. Y. Wu, J. Phys. A {\bf 37}, 6653 (2004)].  By
formulating the problem differently, we obtain a new expression
for the two-point resistance between two arbitrary nodes which is
simpler and can be easier to use in practice. We apply the new
formulation to the cobweb resistor network to obtain the
resistance between  two nodes in the network. Particularly, our
results prove a recently proposed conjecture on the resistance between the center node and a node on the network boundary. Our analysis also solves  the spanning tree problem on the cobweb network.
\end{abstract}

\pacs{01.55+b, 02.10.Yn}

\maketitle

\section{Introduction}
\label{Introduction}

The computation of two-point resistance in a resistor network has a long history.  For a list of relevant references see, e.g., \cite{pol}.  In 2004 one of us \cite{wu2004} derived a compact expression for the two-point resistance in terms of the eigenvalues and eigenvectors of the Laplacian matrix associated with the network. The consideration was soon extended to impedance networks  by Tzeng and Wu \cite{tzengwu} in an analysis making explicit use of the complex nature
of the Laplacian matrix. In practice, however, the use of the result obtained in \cite{wu2004,tzengwu} requires full knowledge of the eigenvalues and eigenvectors of the Laplacian matrix. Due to the fact that the Laplacian is singular,  this task is sometimes difficult to carry through \cite{tan2013}. In this paper we revisit the problem of two-point resistance and derive a new and simpler expression for the resistance. The new expression is then applied to the cobweb resistor network, a problem which has proven to be difficult to analyze  \cite{tan2013}, and the resistance between {\it any} two nodes in the network is obtained. Particularly, our results prove a recently proposed conjecture on the resistance between the center node and a node on the cobweb network boundary \cite{tan2013}. As a byproduct of our analysis, we solve the problem  of spanning trees on the cobweb network.

The organization of this paper is as follows: In Sec. \ref{General} we review the Kirchhoff formulation of a resistance network and outline the derivation of the result of \cite{wu2004}. In Sec. \ref{New} we present a  simpler version of the Kirchhoff formulation which is easier to analyze, obtaining a result different from that reported in \cite{wu2004}. In Sec. \ref{Resistors} the new formulation is applied to the cobweb resistor network obtaining the resistance between any two nodes.  In Sec. \ref{Conjecture} we show our results prove a recent conjecture on the resistance between the center node and a node on the cobweb boundary. Finally in Sec. \ref{Spanning}, we deduce the spanning tree generating function of the cobweb network. A brief summary is given in Sec. \ref{Summary}.

\section{Formulation of two-point resistance}
\label{General}

We first review elements of the theory of two-point resistance.

Let ${\cal L}$ represent a resistor network consisting of ${\cal N}$ nodes numbered $i=1,2,...,{\cal N}$. Let $r_{ij} = r_{ji}$ be the resistance of the resistor connecting nodes $i$ and $j$, hence, the conductance is
\begin{equation}
c_{ij} = r^{-1}_{ij}=c_{ji}.
\label{conductance}
 \end{equation}
Denote the electric potential at the $i$-th node by $V_i$ and the net current flowing {\it into} the network at the $i$-th node by $I_i$.  Since there exist no sinks or sources of current, we have the constraint
\begin{equation}
\sum_{i=1}^{\cal N} I_i=0.
\label{constraint}
\end{equation}
The Kirchhoff law states that
\begin{equation}
\sum_{j=1 \above0pt  j \neq i}^{{\cal N}} c_{ij}(V_i-V_j)=I_i, \qquad i=1,2,...,{\cal N}.
\label{Kirch}
\end{equation}
Explicitly, equation (\ref{Kirch}) reads
\begin{equation}
{\bf L} \vec V = \vec I,
\label{matrixform}
\end{equation}
where
$$
{\bf L}=\left( \begin{array}{ccccccc}
c_1 & -c_{12}& -c_{13}&\ldots &-c_{1{\cal N}} \\
-c_{21} & c_2 &-c_{23}&\ldots &-c_{2{\cal N}}\\
-c_{31} & -c_{32} &c_3&\ldots &-c_{3{\cal N}}\\
\vdots & \vdots &\vdots &\ddots & \vdots \\
-c_{{\cal N}1}&-c_{{\cal N}2}&-c_{{\cal N}3}&\ldots&c_{{\cal N}}
\end{array} \right)
$$
is the Laplacian matrix of  ${\cal L}$ with
\begin{equation}
c_i =\sum_{j=1 \above0pt  j \neq i}^{\cal N} c_{ij},
\end{equation}
and ${\vec V}$ and ${\vec I}$ are  ${\cal N}$-vectors
$$
{\vec V}=\left( \begin{array}{ccccccc}
V_1\\
V_2\\
V_3\\
\vdots \\
V_{\cal N}
\end{array} \right), \qquad
{\vec I}=\left( \begin{array}{ccccccc}
I_1\\
I_2\\
I_3\\
\vdots \\
I_{\cal N}
\end{array} \right).
$$
The Laplacian matrix ${\bf L}$ is also known as the Kirchhoff matrix, or simply the tree matrix; the latter name is derived from the fact that all cofactors of ${\bf L}$ are equal and equal to the spanning tree generating function for ${\cal L}$, a property we shall use in Sec. \ref{Spanning}. Since the sum of all rows of ${\bf L}$ is equal to zero, the matrix ${\bf L}$ is singular and has one eigenvalue $\lambda_1=0$ with corresponding (normalized) eigenvector ${\vec\Psi}_1=\frac 1 {\sqrt{\cal N}} (1, 1, ..., 1)$.

To compute the resistance $R_{\alpha \beta}$ between arbitrary  two nodes $\alpha$ and $\beta$, we connect $\alpha$ and $\beta$ to an external battery and measure the current $I$ going through the battery while no other nodes are connected to external sources. Let the potentials at the two nodes be, respectively, $V_{\alpha}$ and $V_{\beta}$. Then, by Ohm's law, the desired resistance is
\begin{equation}
R_{\alpha \beta} = \frac{V_{\alpha}-V_{\beta}}{I}.
\label{resistance}
\end{equation}
The computation of $R_{\alpha \beta}$ is now reduced to solving Eq. (\ref{Kirch}) for $V_{\alpha}$ and $V_{\beta}$ with the current given by
\begin{equation}
I_i = I (\delta_{i\alpha}-\delta_{i\beta}).
\label{current}
\end{equation}
The solution involves inverting Eq. (\ref{matrixform}) which, unfortunately, cannot be  carried out since ${\bf L}$ is singular. This difficulty is resolved in \cite{wu2004} by considering instead the matrix ${\bf L} (\epsilon) = {\bf L} + \epsilon {\bf I}$, where ${\bf I}$ is the identity matrix, with the parameter $\epsilon$  setting to  zero at the end.

Let the orthonormal eigenvectors of  ${\bf L}$ be ${\vec\Psi}_i  = ( \psi_{i1}, \psi_{i2}, ..., \psi_{i{\cal N}})$, $i=1, 2, ..., {\cal N}$, with eigenvalues $\lambda_i$, namely,
\begin{equation}
{\bf L} \vec\Psi_i = \lambda _i \vec\Psi_i\, ,  \qquad  i=1,2,...{\cal N}.
\label{oldequation}
\end{equation}
Here, as noted earlier, we have one eigenvalue $\lambda_1 = 0$. The above procedure  then gives  the following expression for the two-point resistance \cite{wu2004},
\begin{equation}
R_{\alpha \beta}=\sum_{i=2}^{\cal N}\frac{|\psi_{i\alpha}-\psi_{i\beta}|^2}{\lambda_i}\, ,
\label{resistorold}
\end{equation}
where the summation  is over the  ${\cal N}-1$ nonzero eigenvalues $\lambda_i,\, i = 2, 3,...,{\cal N}$.

\section{New formulation}
\label{New} The formulation of the two-point resistance Eq. (\ref{resistorold}) holds in general. Due to the fact that ${\bf L}$ is singular, however, the actual application of Eq. (\ref{resistorold}) is sometimes difficult to carry through such as in the case of the cobweb network \cite{tan2013}. In this section we derive an alternate and simpler expression for the two-point resistance suitable to networks such as the cobweb.

Under the constraint of Eq. (\ref{constraint}), the sum of the ${\cal N}$ equations in Eq. (\ref{Kirch}) produces the identity $0=0$ so we actually have only ${\cal N}-1$ independent equations in Eq. (\ref{Kirch}). This means we can neglect one redundant equation.  Without the loss of generality we choose to delete the equation numbered $i=1$. Furthermore, we can choose the potential at node $1$ to be $V_1 = 0$. Then the ${\cal N}$ equations in (\ref{Kirch}) and (\ref{matrixform}) reduce to a set of ${\cal N}-1$ equations,
\begin{equation}
\sum_{j=1 \above0pt  j \neq i}^{\cal N} c_{ij}(V_i-V_j)=I_i, \qquad i = 2 ,..., {\cal N}
\label{Kirch1}
\end{equation}
or
\begin{equation}
{\bf \Delta} \vec {\bf {\cal V}} = \vec {\bf {\cal I}}.
\label{matrix1}
\end{equation}
Here
\begin{equation}
{\bf \Delta}=\left( \begin{array}{ccccccc}
c_2 &-c_{23}&\ldots &-c_{2{\cal N}}\\
-c_{32} &c_3&\ldots &-c_{3{\cal N}}\\
\vdots &\vdots &\ddots & \vdots \\
-c_{{\cal N}2}&-c_{{\cal N}3}&\ldots&c_{{\cal N}}
\end{array} \right)
\label{Delta}
\end{equation}
is the $ ({\cal N}-1)\times ({\cal N}-1)$ cofactor of the $\{1,1\}$-element of the Laplacian ${\bf L}$ and
\begin{equation}
{\vec {\cal V}}=\left( \begin{array}{ccccccc}
V_2\\
V_3\\
\vdots \\
V_{\cal N}
\end{array} \right), \qquad
{\vec {\cal I}}=\left( \begin{array}{ccccccc}
I_2\\
I_3\\
\vdots \\
I_{\cal N}
\end{array} \right).
\label{N1vectors}
\end{equation}

Equation (\ref{matrix1}) can now be straightforwardly solved for ${\vec {\cal V}}$ since ${\bf \Delta^{-1}}$ is not singular. Multiplying Eq. (\ref{matrix1}) from the left by ${\bf \Delta}^{-1}$, we obtain the solution ${\cal V}={\bf \Delta}^{-1} {\cal I}$. Explicitly, this reads
\begin{equation}
V_i=\sum_{j=2}^{\cal N} (\Delta^{-1})_{ij} I_j, \qquad i=2,...,N,
\label{Vi}
\end{equation}
where $(\Delta^{-1})_{ij}$  is the $ij$th elements of the inverse matrix ${\bf \Delta^{-1}}$. Combining Eqs. (\ref{resistance}) and (\ref{current}) with Eq. (\ref{Vi}), we obtain the resistance between any two nodes $\alpha$ and $\beta$ other than the node $1$ as
\begin{equation}
R_{\alpha \beta}=(\Delta^{-1})_{\alpha \alpha} + (\Delta^{-1})_{\beta \beta}  - (\Delta^{-1})_{\alpha \beta}- (\Delta^{-1})_{\beta \alpha}.
\label{resistor1}
\end{equation}

Similarly, if one of the nodes, say $\alpha$, is the node $1$ where we have set $V_1 = 0$, Ohm's law gives
\begin{equation}
R_{1\beta} = (\Delta^{-1})_{\beta \beta}.
\label{resistor2}
\end{equation}
Denote by $\vec \Phi_i  = (\phi_{i1}, \phi_{i2},...,\phi_{i{\cal N}} )$ and $\Lambda_i$  the eigenvectors and eigenvalues of ${\bf \Delta}$, namely,
\begin{equation}
{\bf \Delta} \vec \Phi_i = \Lambda_i \vec \Phi_i, \qquad i=2,3,...,{\cal N}.
\label{Delta1}
\end{equation}
Since ${\bf \Delta}$ is Hermitian,
the eigenvectors $\vec \Phi_i $ can be taken to be orthonormal
\begin{equation}
(\vec \Phi_i^{*} ,\,  \vec \Phi_j) =\sum_{\alpha=1}^N
\phi_{i\alpha}^{*}\,\phi_{j\alpha}=\delta_{ij}.\label{ortho}
\end{equation}

Let ${\bf U}$ be the unitary matrix which diagonalizes ${\bf \Delta}$,
\begin{equation}
{\bf U^{\dagger}  \Delta U} = {\bf \Lambda},\nonumber
\label{diagonal}
\end{equation}
where ${\bf \Lambda}$ is diagonal  with elements $\Lambda_i\, \delta_{ij}$. The inverse of Eq. (\ref{diagonal}) is
\begin{equation}
{\bf U^{\dagger} \Delta^{-1} U}={\bf \Lambda}^{-1},
\label{inverse}
\end{equation}
where ${\bf \Lambda}^{-1}$ has elements $\Lambda_i^{-1}\,\delta_{ij}$. It follows that we have
\begin{equation}
{\bf \Delta^{-1}}={\bf U \Lambda^{-1} U^{\dagger}},\nonumber
\end{equation}
or, explicitly,
\begin{eqnarray}
\Delta^{-1}_{ij}&=&\sum_{k=2}^{\cal N}\frac{U_{i k}\,U_{j k}^*}{\Lambda_k}
=\sum_{k=2}^{\cal N}\frac{\phi_{k i}\,\phi_{k j}^*}{\Lambda_k}.
\label{Delta3}
\end{eqnarray}
Substituting Eq. (\ref{Delta3}) into Eq. (\ref{resistor1}) we obtain the expression
\begin{equation}
R_{\alpha \beta}=\sum_{k=2}^{\cal N}\frac{|\phi_{k\alpha}-\phi_{k\beta}|^2}{\Lambda_k}.
\label{resistorfinal}
\end{equation}
Similarly from Eq. (\ref{resistor2}), we have
\begin{equation}
R_{1 \beta}=\sum_{k=2}^{\cal N}\frac{|\phi_{k\beta}|^2}{\Lambda_k}.
\label{resistorfinal1}
\end{equation}

Note the similarity between  Eqs. (\ref{resistorfinal}) and (\ref{resistorold}) in appearance. However, Eq. (\ref{resistorfinal}) can be advantageous since it expresses the resistance $R_{\alpha \beta}$ through the eigenvectors and eigenvalues of the cofactor matrix ${\bf \Delta}$ which is not singular, and the summation  does not
require the singling out of a zero eigenvalue term. The two expressions
(\ref{resistorfinal}) and (\ref{resistorold}) are  different in substance.

\section{The cobweb resistor network}
\label{Resistors}

The cobweb lattice ${\cal L}_{\rm cob}$ is an $M \times N$ rectangular lattice with periodic boundary condition in one direction and nodes on one of the two boundaries in the other direction connected to an external common node.  Therefore there
is a total of $M N+1$ nodes. The example of an $M=3, N=8$ cobweb with resistors $s$ and $r$ in the two directions is shown in Fig. 1. Topologically ${\cal L}_{\rm cob}$ is of the form of a wheel consisting of N spokes and M concentrate circles. There has been considerable recent interest in studying the resistance in a cobweb network (for a summary of related works, see \cite{tan2013}). But there has been no generally valid exact result.
\begin{figure}
\epsfxsize=90mm \vbox to2in{\rule{0pt}{2in}}
\includegraphics{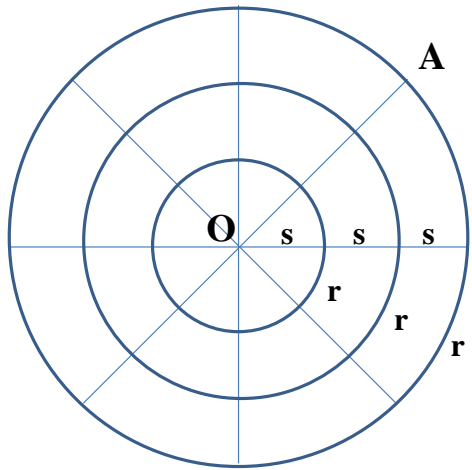}  \caption{$M \times N$ cobweb network with M=3
and N=8. Bonds in spokes and circular directions are  resistors
$s$ and $r$. The center is denoted by $O$ and $A$ denotes a point
on the boundary.}\label{fig1}
\end{figure}
%\begin{figure}[tbp]
  % \includegraphics[width=0.35\textwidth]{Fig1Cobweb.eps}
 %  \includegraphics[width=0.3\textwidth]{Fig1Cobweb.eps}
  %   \caption{$M \times N$ cobweb network with M=3 and N=8. Bonds in spokes and
   %   circular directions are  resistors $s$ and $r$. The center is denoted by $O$ and
    %   $A$ denotes a point on the boundary.} \label{fig1}
     %  \end{figure}
To compute resistances on the cobweb network, we make use of the formulation given in the preceeding section, and choose the center node $O$ to be the node $1$ in the cobweb Laplacian ${\bf L}_{\rm cob}$. This leads us to consider the $(MN) \times (MN)$ cofactor of the $\{1,1\}$-element of ${\bf L}_{\rm cob}$, namely,
\begin{equation}
{\bf \Delta}_{M  N}=r^{-1}{\bf L}_N^{\rm{per}}\otimes {\bf I}_M+s^{-1}{\bf I}_N \otimes {\bf L}_M^{(DN)},
\label{MtimesN}
\end{equation}
where ${\bf L}_N^{\rm{per}}$ can be thought of as the Laplacian of a 1D lattice with periodic boundary conditions,
$$
{\bf L}_N^{\rm{per}}=\left( \begin{array}{ccccccc}
2 & -1 & 0 &\ldots &0&0&-1 \\
-1 & 2 & -1&\ldots &0 &0&0\\
\vdots & \vdots & \vdots &\ddots &\vdots & \vdots & \vdots \\
0&0&0&\ldots&-1&2&-1\\
-1&0&0&\ldots&0&-1&2
\end{array} \right),
$$
and ${\bf L}_M^{(DN)}$ the Laplacian of a 1D lattice with Dirichlet-Neumann boundary conditions,
\begin{equation}
{\bf L}_M^{(DN)}=\left( \begin{array}{ccccccc}
2 & -1 & 0 &\ldots &0&0&0 \\
-1 & 2 & -1&\ldots &0 &0&0\\
\vdots & \vdots & \vdots &\ddots &\vdots & \vdots & \vdots \\
0&0&0&\ldots&-1&2&-1\\
0&0&0&\ldots&0&-1&1   \end{array} \right) \nonumber
\end{equation}
Here, ${\bf I}_M$ and ${\bf I}_N$ are identity matrices.

The eigenvalues and eigenvectors of ${\bf L}_N^{\rm per}$ and ${\bf L}_M^{(DN)}$ are known to be, respectively,
\begin{eqnarray}
f_n(x) &=& \sqrt{{1}/{N}}\; {\rm exp}(2\,  \theta_n \, x ), \nonumber\\
\Lambda_n &=& 2 - 2\cos (2 \, \theta_n) , \qquad n=0,1,...,N-1,
\nonumber
\end{eqnarray}
and
\begin{eqnarray}
f_m(y) &=& \frac{2}{\sqrt{2M+1}}\sin (2 \, \varphi_m )\, , \nonumber\\
\Lambda_m &=& 2 - 2\cos (2 \varphi_m) , \qquad m=0,1,...,M-1,
\label{lambda}
\end{eqnarray}
where
\begin{equation}
\theta_n=\frac{n\,\pi}{N}, \qquad  \qquad \varphi_{m}=\frac{(m + \frac 1 2) \pi}{2M+1}.
\label{phi}
\end{equation}
This leads to the following eigenvalues and eigenvectors
for  the cofactor matrix
${\bf \Delta}_{M N}$,
\begin{eqnarray}
\Lambda_{m,n}&=&2r^{-1}(1-\cos{2\theta_n})+2s^{-1}(1-\cos{2\varphi_{m}}),\nonumber \\
\phi_{(m,n);(x,y)}&=&\frac{2}{\sqrt{N(2M+1)}}\exp{(2 i x \theta_n)}\sin{(2y\varphi_{m}}). \label{psi}
\end{eqnarray}
Therefore using
Eq. (\ref{resistorfinal}), the resistance $R^{\rm{cob}}(r_1,r_2)$ between
two nodes at $r_1 = \{x_1, y_1\}$ and $r_2 = \{x_2, y_2\}$, when both not the center $O$, is
\begin{eqnarray}
R ^{\rm{cob}}(r_1,r_2)&=&\sum_{m=0}^{M-1}\sum_{n=0}^{N-1}
\frac{\left|\phi_{(m,n);(x_1,y_1)}-\phi_{(m,n);(x_2,y_2)}\right|^2}
{\Lambda_{m,n}}\nonumber\\
&=&\frac{2r}{N(2M+1)}\sum_{m=0}^{M-1}\sum_{n=0}^{N-1}
\frac{S_1^2+S_2^2-2S_1S_2\cos[2(x_1-x_2)\theta_n]}
{(1-\cos{2\theta_n})+ h (1-\cos{2\varphi_m})}
\label{R12}
\end{eqnarray}
where
\begin{equation}
h=r/s, \qquad S_1=\sin\left(2y_1\varphi_{m}\right), \qquad S_2=\sin\left(2y_2\varphi_{m}\right). \nonumber
\end{equation}

Introduce $\varLambda_m =\varLambda(\varphi_m)$ by writing
\begin{equation}
1 + h (1-\cos \varphi_m ) = \cosh 2 \varLambda_m \nonumber
\end{equation}
or, equivalently,
\begin{equation}
\sinh \varLambda_m = \sqrt h \sin \varphi_m\, .
\end{equation}
We can then carry out the summation over $n$ in (\ref{R12}) by using the summation identities \cite{note}
\begin{equation}
\frac 1 N \sum_{n=0}^{N-1} \frac {\cos (2\, \ell\, \theta_n)}
 {\cosh 2 \varLambda - \cos 2\theta_n   } = \, \frac{\cosh[(N -2 \, \ell)\varLambda)]}
{\sinh (2\varLambda )\sinh(N \varLambda)}, \quad {\rm with}  \quad \ell = 0, \> |x_1-x_2|\, .
\label{sumidentity}
\end{equation}
 to obtain
\begin{eqnarray}
R ^{\rm{cob}}(r_1,r_2)&=& \frac{2r} {2M+1}
\sum_{m=0}^{M-1}\frac{S_1^2+S_2^2-2S_1 S_2
\cosh\big[2|x_1-x_2|\,\varLambda_m\big]}
{\sinh(2\,\varLambda_m)}\,\coth ( N\,\varLambda_m ) \nonumber\\
&+&\frac{2r} {2M+1}  \sum_{m=0}^{M-1}
\frac{ 2S_1S_2\sinh\big[2|x_1-x_2|\,\varLambda_m \big]}
{\sinh (2\,\varLambda_m)}. \label{R12y}
\end{eqnarray}
%where $\varLambda_m = \varLambda (\varphi_m)$.

In the special case of $x_1 = x_2=x$, i.e., two nodes in the same $y$ column at $y_1$ and $y_2$,
Eq. (\ref{R12y}) reduces to
\begin{equation}
R ^{\rm{cob}}(\{x,y_1\}, \{x, y_2\}) = \frac{2r} {2M+1}\sum_{m=0}^{M-1}\frac {\coth (N \varLambda_m)}
  {\sinh (2\varLambda_m)} \big[\sin (2y_1 \varphi_m) -\sin (2y_2 \varphi_m)\big]^2,
\label{x1x2}
\end{equation}
and in the special case of $ y_1 = y_2=y$, i.e., two nodes in the same $x$ row at $x_1$ and $x_2$,
Eq. (\ref{R12y}) reduces to
\begin{equation}
R ^{\rm{cob}}(\{x_1,y\}, \{x_2, y\}) = \frac{8r} {2M+1}\sum_{m=0}^{M-1}
\frac { \sinh \big[|x_1-x_2| \varLambda_m\big]
        \sinh \big[ \big(N-|x_1-x_2| \big) \varLambda_m \big]  }
      {\sinh (2 \varLambda_m ) \sinh (N \varLambda_m) }  \sin^2 (2 y \varphi_m). \label{y1y2}
\end{equation}
 Note that the result (\ref{x1x2}) is independent of the position $x$ as it should.

If one of the two nodes is the center $O$ of the cobweb and the
other node at $P = \{x,y\}$, then we use
Eq. (\ref{resistorfinal1}) and obtain the resistance
\begin{eqnarray}
R^{\rm{cob}}(O,P)&=& \sum_{m=0}^{M-1}\sum_{n=0}^{N-1}
\frac {   |\phi_{(m,n);(x,y )} | ^2   }
{ \Lambda_{m,n} } \nonumber \\
&=&
\frac{2r}{N(2M+1)} \sum_{m=0}^{M-1} \sum_{n=0}^{N-1}
\frac {\sin^2 ( 2y\varphi _m )}
{(1-\cos2 \theta_n )+h (1-\cos2 \varphi_m )} \nonumber \\
&=&  \frac{2r} {2M+1}  \sum_{m=0}^{M-1}
\frac { \coth ( N\,\varLambda_m )} {\sinh
(2\,\varLambda_m)} \, {\sin^2  (2\, y \, \varphi_m)} , \quad y = 1,2,...,M .
\label{ROAy}
 \end{eqnarray}
 Note that the result (\ref{ROAy}) is independent of the position $x$ as it should.

In the special case of the resistance between the center $O$ and a
point $A=\{x,N\}$ on the outer boundary of the cobweb, we use $y=M$ and obtain from (\ref{ROAy})
\begin{equation}
R^{\rm cob}(O, A)= \frac{2r}{2M+1}\sum_{m=0}^{M-1}
\frac { \coth ( N\,\varLambda_m )} {\sinh
(2\,\varLambda_m)}\, {\cos^2 \varphi_m},
\label{ROAfinal}
\end{equation}
where use has been made of the identity
\begin{equation}
\sin ( 2\,M\,\varphi_m) = (-1)^{m} \cos \varphi_m, \nonumber
\end{equation}
 which is a consequence of the fact  $2M \varphi_m + \varphi_m  = \big( m+\frac 1 2 \big) \pi$.

In the limit of $N\to \infty$, we replace $\coth ( N\,\varLambda_m ) \to 1$ in (\ref{R12y}), (\ref{x1x2}),
(\ref{ROAy}) and
(\ref{ROAfinal}), and replace $\sinh \big[ \big(N-|x_1-x_2| \big) \varLambda_m \big]  /
    \sinh (N \varLambda_m) \to e^{-|x_1-x_2|\varLambda_m }$ in (\ref{y1y2}).

In the limit of  $M\to\infty$, we convert the summations in
(\ref{R12y}) - (\ref{ROAfinal}) into integrals by making
use of the replacement
\begin{equation}
\frac{1}{2M+1}\sum_{m=0}^{M-1} F(\varphi_m) \to \frac 1 \pi
\int_0^{\pi/2} F(\varphi)\, d \varphi , \nonumber
\end{equation}
which is an identity valid for  any function $F(\varphi_m)$.

Equations (\ref{R12y}) - (\ref{ROAfinal})
are our main results for the cobweb resistor
network.

\section{Proof of the TZY conjecture}
\label{Conjecture}
In this section we prove a recent conjecture on  $R^{\rm cob}(O,A)$ due to Tan, Zhou and Yang \cite{tan2013}, the TZY conjecture. The TZY conjecture was also cited in
\cite{tan13Sept} in an analysis of the $4\times N$ cobweb network.

Using previous known results for $M=1,2$ and algebraic results
for $M=3$ obtained after elaborate algebraic calculations, Tan, Zhou and Yang \cite{tan2013} conjectured that the resistance between the
center node $O$ and a node $A$ on the boundary of an $M\times N$  cobweb is
\begin{equation}
R^{\rm{cob}}(O,A)=r\sum_{m=0}^{M-1} \frac{2+p_m}{2M+1} \cdot \frac{{\rm
coth} (N\ln{\sqrt{T_m}} ) } {T_m-T_m^{-1}} \qquad ({\rm TZY\>\>conjecture})
 \label{ROAtan}
\end{equation}
where
\begin{eqnarray}
p_m&=&2\cos (2\, \varphi_m) ,\nonumber\\
 T_m&=&1+ h -\frac{ h\,p_m}{2}+\sqrt{\left(1+ h -\frac{\, h p_m}{2}\right)^2-1}\,  . \nonumber
 \end{eqnarray}
Here $\varphi_m = (m+\frac 1 2)\pi/(2M+1)$ as defined in (\ref{phi}),
and the summation in (\ref{ROAtan}) is taken over $m=0, 1, ..., M-1$ (as versus $m=1,2,...,M$ in \cite{tan2013}).

Now, we have the identities
\begin{eqnarray}
\cosh^{-1} z &=& \ln (z + \sqrt {z^2 -1}) \nonumber\\
  \cosh^{-1} (1+h-h\cos 2 z ) &=& 2\, \sinh^{-1} (\sqrt h \sin z) =\ \  2 \, \varLambda (z). \label{iden}
\end{eqnarray}
 It is then easy using the identities (\ref{iden}) to see that we have
 \begin{eqnarray}
\ln \sqrt {T_m} &=& \varLambda (\varphi_m  ) \nonumber \\
     T_m - {T_m}^{-1} &=& 2 \sinh \big[2 \, \varLambda (\varphi_m  )\big].
\label{Tm}
\end{eqnarray}
Substituting  (\ref{Tm}) and $\ 2+p_m = 4 \cos^2 \varphi_m \ $ into (\ref{ROAtan}),
the TZY conjecture  (\ref{ROAtan}) reduces to our exact result (\ref{ROAfinal}).

\section{Spanning tree on Cobweb network}
\label{Spanning}
As a byproduct of our analysis, we solve the problem of enumerating weighted spanning trees on
an $M\times N$  cobweb network ${\cal L}_{\rm{cob}\ M\times N}$.

The problem of enumerating spanning trees on a graph was first considered by Kirchhoff \cite{Kirchhoff} in his analysis of electrical networks. The enumeration of spanning trees concerns the evaluation of the tree generating function
\begin{equation}
Z_{\rm{cob}\ (M\times N)}^{\rm Sp}(x,y)=\sum_T x^{n_x}y^{n_y}
\label{span}
\end{equation}
where we assign weights $x$ and $y$, respectively, to edges in the spokes and circle directions, and the summation is taken over all spanning tree configurations T on
${\cal L}_{ {\rm{cob}}\ (M\times N)  }$
with $n_x$ and $n_y$ edges  in the respective directions. Setting $x=y=1$ we obtain
\begin{equation}
Z_{\rm{cob}\ (M\times N)}^{\rm Sp}(1,1)=\mbox{the number of spanning trees on cobweb network}.
\label{Nspan}
\end{equation}

It is well-known \cite{Brooks,Harary,tzengwu1} that the spanning tree generating function is given by the determinant of the cofactor of {\it any} element of the Laplacian matrix of the network.
We can therefore  evaluate ${\bf \Delta}_{MN}$ given in (\ref{MtimesN}) with $r^{-1}=x, s^{-1}=y$. This gives
\begin{eqnarray}
Z_{\rm{cob}\ (M\times N)}^{\rm Sp}(x,y) &=& \det |{\bf \Delta}_{MN}| \nonumber \\
                         &=& \prod_{m=0}^{M-1}\prod_{n=0}^{N-1}\Lambda_{m,n}(x,y),
\label{span2}
\end{eqnarray}
where $\Lambda_{m,n}(x,y)$ is given by Eq. (\ref{psi}) with $r^{-1}= x$ and $s^{-1} =y$. Thus, we obtain the closed form expression for the spanning tree generating function
\begin{eqnarray}
 Z_{\rm{cob}\ (M\times N)}^{\rm Sp}(x,y) &=& \prod_{m=0}^{M-1}\prod_{n=0}^{N-1}\left[2x\left(1-\cos{\frac{2\pi n}{N}}\right)+2y\left(1-\cos{\frac{\pi(2m+1}{2M+1}}\right)\right]
\nonumber \\
&=&\prod_{m=0}^{M-1}\prod_{n=0}^{N-1}4\left[x \sin^2{\frac{\pi
n}{N}}+y\sin^2{\frac{\pi(m+\frac 1 2 )}{2M+1}}\right]\label{spangen2}.
\end{eqnarray}

In comparison, the spanning tree generating function for an $M \times N$ cylindrical lattice periodic in the $N$ or $x$ direction computed by Tzeng and Wu \cite{tzengwu1} is
 \begin{equation}
Z_{\rm{cyl}\ M\times N}^{\rm Sp}(x,y) = \frac{1}{M N} \prod_{m=0}^{M-1}
\prod_{n=0 \above0pt  (m,n) \neq (0,0)}^{N-1} \left[2x\left(1-\cos{\frac{2 n
\pi}{N}}\right)+2y\left(1-\cos{\frac{m
\pi}{M}}\right)\right].
\label{spancyl1}
 \end{equation}
The expression (\ref{spancyl1}) can be transformed to
\begin{equation}
Z_{\rm{cyl}\ M\times N}^{\rm Sp}(x,y)=N x^{N-1}y^{M-1}
\prod_{m=1}^{M-1}\prod_{n=1}^{N-1} 4\left[x \sin^2{\frac{\pi
n}{N}}+y\sin^2{\frac{\pi m}{2M}}\right]\label{spancylfin}
\end{equation}
by using the identities
\begin{equation}
\prod_{n=1}^{N-1} 4 x \sin^2 \frac {\pi n} N  = N^2 x^{N-1} , \qquad
\prod_{m-1}^{M-1} 4 y \sin^2 \frac {\pi m} { 2 M} = M\,  y^{M-1}  .\nonumber
\end{equation}
The expression (\ref{spancylfin}) can now be compared to (\ref{spangen2}) for the $M\times N$ cobweb.
Particularly, for $M=3, N=8$,
we obtain for the $3\times 8$ cobweb the number
 \begin{equation}
 Z_{\rm{cob}\ (3\times 8)}^{\rm Sp}(1,1)= 167\ 999\ 155\ 129, \nonumber
\end{equation}
and for the  $3\times 8$ cylinder the number
\begin{equation}
Z_{\rm{cyl}\ (3\times 8)}^{\rm Sp}(1,1)=1\ 633\ 023\ 000. \nonumber
\end{equation}
 The addition of one center node to a $3\times 8$ cylinder increases the number of spanning trees
by more than 100 times!

Finally, since both the cobweb and cylindrical lattices are the rectangular lattice  with
different boundary conditions which do not affect the bulk limit, they have the same growth constant,
or spanning tree constant as given in \cite{Temperley72,Wu77},
\begin{eqnarray}
z &=& \lim_{M,N\to \infty} (MN)^{-1} \ln Z _{(M\times N)} (1,1)\nonumber \\
  &=& \frac 4 \pi ( 1 -  3^{-2} + 5^{-2} - 7^{-2} + \cdots ) = 1.166\ 243\ 6 \dots \ \ . \nonumber
\end{eqnarray}

\section{Summary and Discussions}
\label{Summary}
We have re-visited the problem of the evaluation of two-point resistances
in a resistor network $\cal L$ considered in \cite{wu2004}, and
re-formulated the evaluation in terms of the eigenvalues and eigenfunctions of a cofactor of
the Laplacian of $\cal L$.  The new formulation is applied to the cobweb resistor network,
a cylindrical lattice with sites on one cylinder boundary connected to an external common center site $O$
as shown in Fig. \ref{fig1}, which has heretofore
eluded exact analysis.  Our analysis leads to exact expressions (\ref{R12y}), (\ref{ROAy}) and (\ref{ROAfinal}),
respectively,
for the resistance between arbitrary two nodes on the cylinder, between
the center $O$ and any other  point $P$ on the cylinder,
 and between the center $O$ and a
point $A$ on the open cylinder boundary.  Particularly, the result (\ref{ROAfinal})
trivially verifies a conjecture
by Tan, Zhou and Yang \cite{tan2013}.
 We also obtain the generating function  (\ref{spangen2}) of spanning trees on the cobweb lattice.

Finally, we remark that our results on cobweb resistor networks
also apply to cobweb capacitance networks \cite{tan13Sept} such as the one shown in FIG. 1 with
capacitances $C$ and $C_0$ in place of $r$ and $s$.
Our analysis goes through with the replacement of $r,s$ by $1/C, 1/C_0$, respectively.

\section{Acknowledgment}
\label{Acknowledgment}

The work of N.Sh.I. and R.K. was supported by a Marie Curie IIF (Project No. 300206 - RAVEN) and IRSES (Project No. 295302 - SPIDER) within 7th European Community Framework Programme and by the grant of the Science Committee of the Ministry of Science and Education of the Republic of Armenia under contract 13-1C080. We thank Professor Z.-Z. Tan for sending a copy of Ref. \cite{tan13Sept} prior to publication.

\end{document}